\documentclass[conference]{IEEEtran}
\IEEEoverridecommandlockouts
\usepackage{cite}
\usepackage{amsmath,amssymb,amsfonts}
\usepackage{algorithmic}
\usepackage{graphicx}
\usepackage{textcomp}
\usepackage{xcolor}
\usepackage{subfig}
\usepackage{chemarrow}
\usepackage{bbding}
\usepackage{multirow}
\usepackage{makecell}
\usepackage{diagbox}

\usepackage{hyperref}
\usepackage{array}
\def\BibTeX{{\rm B\kern-.05em{\sc i\kern-.025em b}\kern-.08em
    T\kern-.1667em\lower.7ex\hbox{E}\kern-.125emX}}
\usepackage{bbold}
\begin{document}

\title{Informing selection of performance metrics for medical image segmentation evaluation using configurable synthetic errors\\
}

\author{\IEEEauthorblockN{Shuyue Guan$^1$, Ravi K. Samala, Weijie Chen}
\IEEEauthorblockA{Division of Imaging, Diagnostics, and Software Reliability \\
Office of Science and Engineering Laboratories, Center for Devices, and Radiologic Health\\
United States Food and Drug Administration, Silver Spring, Maryland, United States \\
\{shuyue.guan, ravi.samala, weijie.chen\}@fda.hhs.gov \\ $^1$\url{https://orcid.org/0000-0002-3779-9368}
}
}
\IEEEoverridecommandlockouts

\maketitle

\IEEEpubidadjcol

\begin{abstract}
Machine learning-based segmentation in medical imaging is widely used in clinical applications from diagnostics to radiotherapy treatment planning. Segmented medical images with ground truth is useful to investigate the properties of different segmentation performance metrics to inform metric selection. Regular geometrical shapes are often used to synthesize segmentation errors and illustrate properties of performance metrics, but they lack the complexity of anatomical variations in real images. In this study, we present a tool to emulate segmentations by adjusting the reference (truth) masks of anatomical objects extracted from real medical images. Our tool is designed to modify the defined truth contours and emulate different types of segmentation errors with a set of user-configurable parameters. We defined the ground truth objects from 230 patient images in the Glioma Image Segmentation for Radiotherapy (GLIS-RT) database. For each object, we used our segmentation synthesis tool to synthesize 10 versions of segmentation (\textit{i.e.}, 10 simulated segmentors or algorithms) where each version has a pre-defined combination of segmentation errors. We then applied 20 performance metrics to evaluate all synthetic segmentations. We demonstrated the properties of these metrics including their ability to capture specific types of segmentation errors. By analyzing the intrinsic properties of these metrics and categorizing the segmentation errors, we are working toward the goal of developing a decision-tree tool for assisting selection of segmentation performance metrics.
\end{abstract}


\section{Introduction}
Image segmentation techniques are widely used to identify the extent of anatomical objects (\textit{e.g.}, lesions or organs) in medical images for many clinical applications, such as tumor size monitoring and feature extraction for diagnostics. Objective assessment of segmentation performance plays a key role in ensuring the efficacy of a variety of medical devices involving machine learning-based segmentation algorithms~\cite{Hesamian2019Deep}; however, there are still substantial variations in the assessment methodologies for image segmentation. For example, there is no consensus about how to establish a reference standard using human experts~\cite{Warfield2004STAPLE, Zheng2017Truth}. In addition, there are many metrics that can be used to evaluate segmentation performance, but we do not have the consensus in methods to select performance metrics that are most appropriate for specific clinical tasks~\cite{Taha2014selecting, Taha2015Metrics}. Selection of metrics for evaluating image segmentation algorithms is not trivial due to variabilities in shape, size, and boundary complexity of the regions of interest (ROIs) to be segmented, and the effects of segmentation errors on different clinical tasks.

Segmented medical images with ground truth are useful to investigate the properties of segmentation performance metrics and inform metric selection. However, segmentation ground truth is not available for real medical images. Simple geometrical shapes such as rectangles and ellipses are often used as ground truth objects to illustrate some properties of performance metrics with synthetic segmentation errors~\cite{Taha2015Metrics, Kim2015Quantitative, Nai2021Comparison}, but simple  geometrical shapes lack the complexity of anatomical variations in real images. Therefore, the purpose of this study is to develop a tool to emulate segmentations by adjusting truth masks of anatomical objects extracted from real medical images. Our tool involves a set of user-configurable parameters to modify pre-defined truth contours to emulate different types of segmentation errors (\textit{e.g.}, under/over segmentation and boundary spiculations) thereby generating realistic synthetic segmentations. We then applied segmentation performance metrics to evaluate all synthetic segmentations. We demonstrated the properties of these metrics including their ability to capture specific types of segmentation errors.

\begin{figure*}[tbp]
    \centering
    \includegraphics[width=0.5\textwidth]{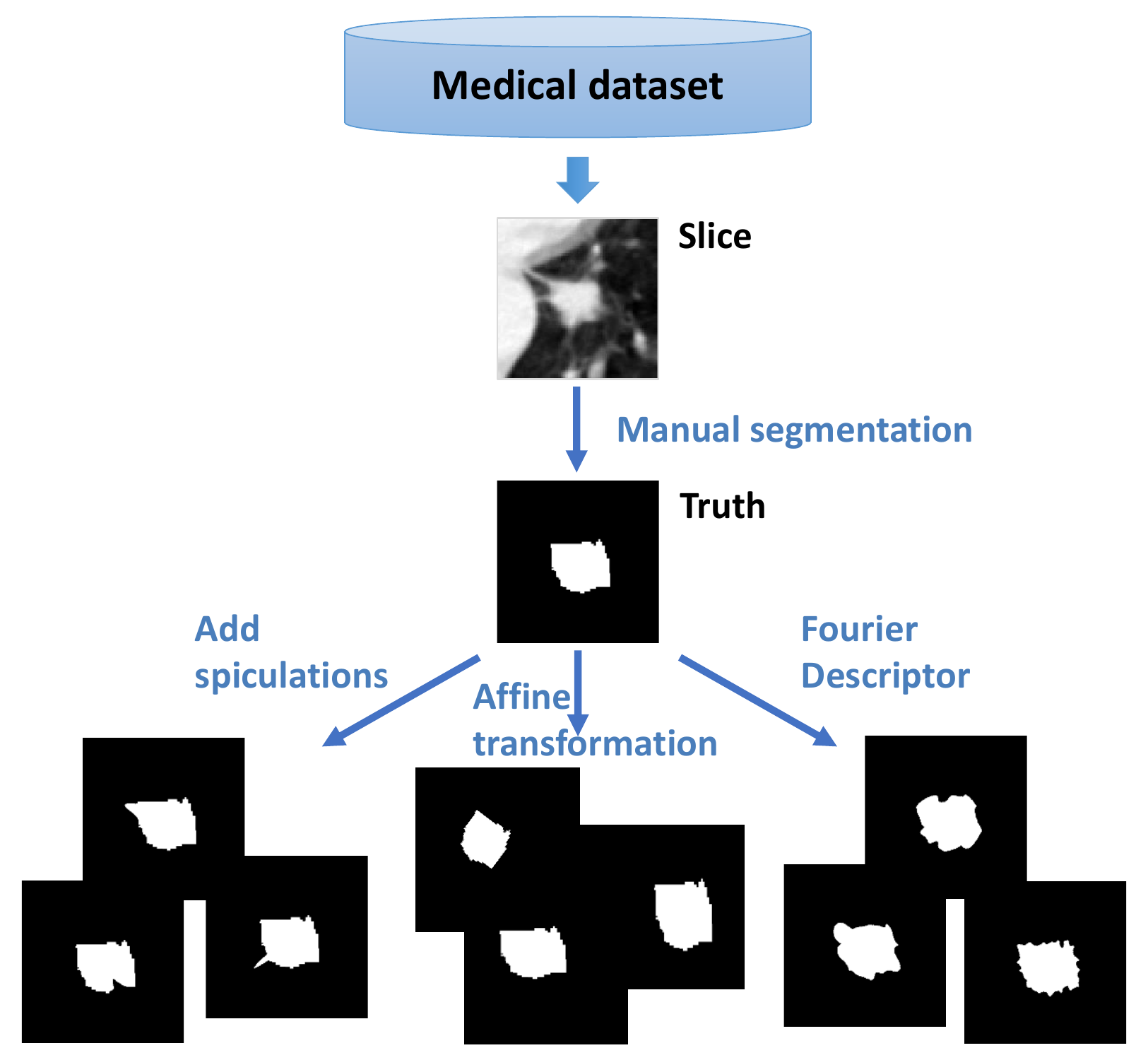}
    \caption{Generation of synthetic medical segmentations: methods overview.}
    \label{fig:methods_overview}
\end{figure*}

\section{Methods}
We developed a Medical Image Segmentation Synthesis tool (MISS-tool) to generate synthetic segmentations based on pre-defined truth masks of objects extracted from real-world medical images. This tool allows users to customize segmentation errors with configurable parameters. The MISS-tool has a graphical user interface (GUI) for tuning parameters and visualizing outcomes. The description of the methods below assumes 2-D images and 3-D images can be processed in a slice-by-slice fashion.

First, masks of objects, \textit{e.g.}, lesions, organs, etc., from real-world medical images are defined as ground truth. These can be manual segmentations by domain experts. If an object has multiple manual segmentations, they can be combined using some truthing methods such as the \textit{Simultaneous Truth And Performance Level Estimation} (STAPLE)~\cite{Warfield2004STAPLE}.
For each object, we emulated segmentations with different types of segmentation errors by using a combination of three methods (Figure~\ref{fig:methods_overview}):
\begin{itemize}
    \item \textbf{Affine transformation:} resizing, shift, rotation, \textit{etc}.
    \item \textbf{Adding spiculation:} adjust contours in the polar coordinate system
    \item \textbf{Fourier Descriptor (FD):} add, remove, or change components in frequency domain
\end{itemize}

\begin{figure*}[tbp]
    \centering
    \includegraphics[width=0.8\textwidth]{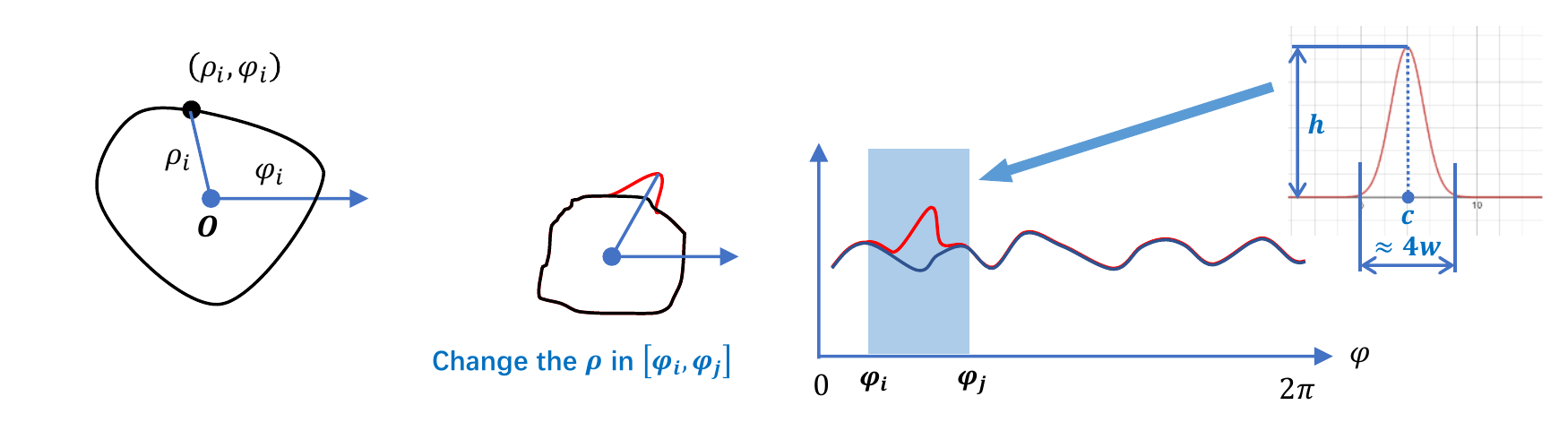}
    \caption{Contour adjustment in polar coordinate system.}
    \label{fig:add_sp}
\end{figure*}

\subsection{Affine transformation}
An emulated segmentation can be obtained by modifying the truth contour with affine transformations. Our MISS-tool provides three types of affine transformation: resizing, shifting, and rotation. Resizing has two parameters to control the size ratio in the horizontal and vertical directions respectively. Shifting has two parameters characterizing the uniform change of $(x, y)$ coordinates of all pixels in the truth mask. Rotation is defined with one angular parameter. 

\subsection{Boundary spiculation}
Changing boundary spiculations is used to emulate this specific type of segmentation errors on the contour. This is achieved by representing each point on the contour in a polar coordinate system, where the coordinate of each point is represented by the distances ($\rho$) and angles ($\varphi$) between its location and the center of mass of the points on the contour. As shown in Figure~\ref{fig:add_sp}, point $i$ on the contour is $(\rho_i, \varphi_i)$ in the polar coordinate system. The contour can be represented by: $\rho(\varphi)$. We modify the contour by adding a Gaussian function (Equation~\ref{eq:gaussian_sp}) to emulate boundary spiculation errors:

\begin{equation}
\label{eq:gaussian_sp}
    G(\varphi)=he^{-(\frac{\varphi-c}{w})^2}
\end{equation}

The function contains three parameters, center ($c$), height ($h$), and width ($w$), to control the location and magnitude of each spiculation. The height can be a positive or negative value to synthesize outward or inward spiculation. The center and width are angle degrees ranging in $[0, 2\pi]$, and as shown in Figure~\ref{fig:add_sp}, the width of the spiculation (Gaussian function) is about four times of the width value ($w$).
The contour after adding a spiculation is: 
\begin{equation*}
    \rho' (\varphi)=\rho(\varphi)+G(\varphi)
\end{equation*}

\begin{figure}[h]
    \centering
    \includegraphics[width=0.4\textwidth]{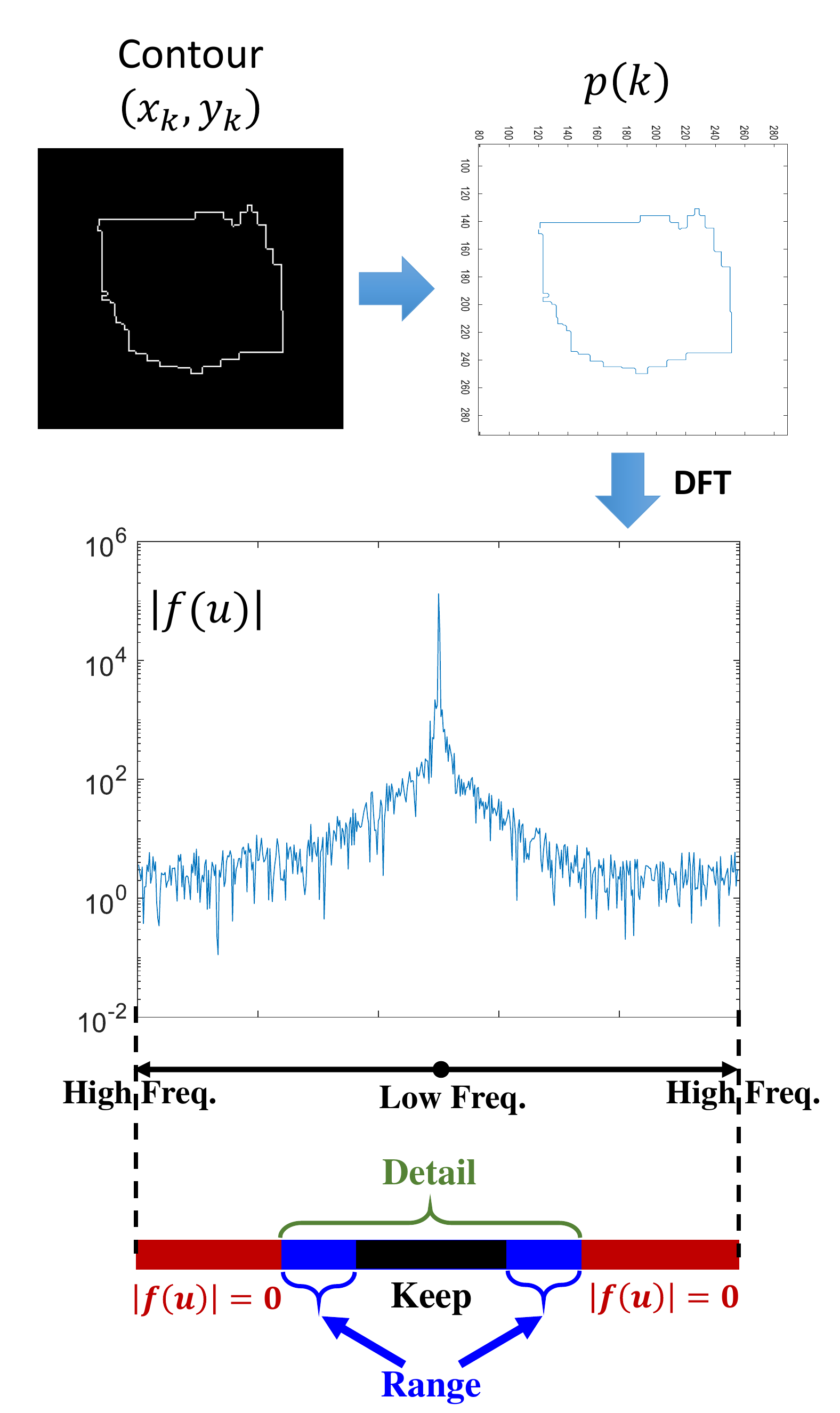}
    \caption{An example of converting the contour points to Fourier descriptors. The plot displays the magnitude of Fourier descriptors re-ordered by frequency. The \textit{Direct Current} (DC)-term: $f(0)$ is in the center; from center to the left and right are low-frequency components to high-frequency components, which are are $f(n-1), f(N-2), \cdots, f(N/2)$ and $f(1), f(2), \cdots, f(N/2)$. }
    \label{fig:fd}
\end{figure}

\subsection{Fourier Descriptor (FD)}
\label{sec:fd}
FD is used to emulate the level of detail in the segmentation contour. The coordinate of each point $k$ on the contour $(x_k,y_k )$ in the Cartesian coordinate system can be written as a complex number:
\[
p(k)=x_k+y_k i
\]
Where $k=0,1,2,\cdots, N-1$, and $N$ is the number of points (pixels) on the contour. By using Discrete Fourier Transform (DFT):
\[
f(u)=\frac{1}{N}\sum_{k=0}^{N-1} p(k) e^{-i \frac{2\pi}{N} uk}
\]
Where $f(u)$ is a Fourier descriptor (FD) of the contour, and $u=0,1,2,\cdots, N-1$. We transform the truth contour into the Fourier domain, where the contour detail is modified to emulate a segmentation. The segmentation contour $p(k)$ can be obtained from the modified FD by applying inverse-DFT:
\[
p(k) \autorightleftharpoons{DFT}{i-DFT} f(u)
\]

Figure~\ref{fig:fd} shows an example of FD representation of a contour. Each descriptor $f(u)$ is a complex number:
\begin{equation}
\label{eq:fd}
f(u)=\alpha_u+\beta_u i    
\end{equation}
The plot in Figure~\ref{fig:fd} displays their magnitude:
\[
|f(u)|=|\alpha_u+\beta_u i| = \sqrt{\alpha^2_u+\beta^2_u}
\]
 As shown in Figure~\ref{fig:fd}, in our MISS-tool, we set the creation of the segmentation contour to keep low-frequency FDs of the truth contour and modify a range of middle-frequency FDs to emualate segmentation errors, and remove high-frequency FDs (set to be zero). The specific range of "middle frequency" is tunable. 

In detail, we provide three parameters to control the generation of a segmentation contour from the truth contour. As shown in the Figure~\ref{fig:fd}, the first parameter is \texttt{Detail}, which is the number of non-zero FDs, and other FDs (red areas in Figure~\ref{fig:fd}) are set to be $|f(u)|=0$. The second parameter is \texttt{Range}, which is the number of FDs to be changed (blue areas in Figure~\ref{fig:fd}). The remaining FDs keep unchanged (black area in Figure~\ref{fig:fd}). The third parameter is \texttt{Magnitude} to control how the specified middle FDs (blue) are modified. In specific, the real part $\alpha_u$ and imaginary part $\beta_u$ in a to-be-modified FD $f(u)$ are changed by:
\[
 \left\{ \begin{array}{l}
\alpha_u' = \alpha_u + r_u m \\
\beta_u' = \beta_u + s_u m 
\end{array}\right.
\]
Where updated $f'(u)=\alpha_u'+\beta_u' i$; $r_u, s_u\in (-0.5,0.5)$ are two uniform random variables; $m$ is the \texttt{Magnitude} parameter.

\section{Materials}

\subsection{Medical image data}
We used images from the medical images (the Glioma Image Segmentation for Radiotherapy (GLIS-RT) dataset~\cite{Gliomadata, Shusharina2021Cross}) to generate synthetic segmentations. The GLIS-RT dataset has CT and MRI head images from 230 patients. The CT images are manually segmented by a radiation oncologist including gross tumor volume (GTV) targets, clinical target volume (CTV) targets, barriers to cancer spread (ventricles, sinuses, cerebellum, \textit{etc}.), and organs at risk (brainstem, optic nerves, eyes, cochleae, \textit{etc}.).

\begin{figure}[h]
    \centering
    \includegraphics[width=0.3\textwidth]{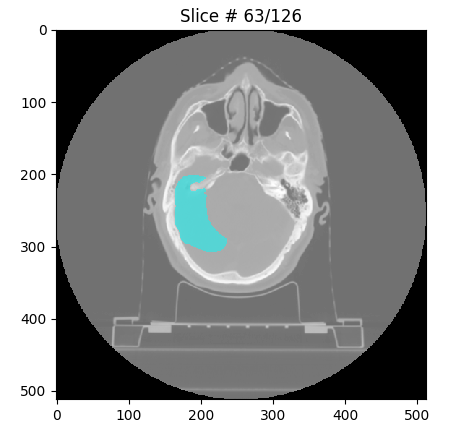}
    \caption{The middle slice in CT images of the GTV target (patient No. GLI\_001\_GBM) overlapped by its delineation (in teal color).}
    \label{fig:gtv_exp}
\end{figure}

We used the DICOM-RT Tool\footnote{\url{https://github.com/brianmanderson/Dicom_RT_and_Images_to_Mask}} in Python to extract the CT images and their delineations. For each patient, only one middle slice in the delineations of GTV targets was extracted and saved (Figure~\ref{fig:gtv_exp}). Thus, we collected 230 (2-D) segmentations of GTV target in total. These segmentations are defined as ground truth for the 230 patients for the purpose of segmentation error synthesis.

\subsection{Synthetic segmentation}
For every GTV target mask (truth), we synthesized 10 versions of segmentation with pre-defined segmentation-error characteristics regarding the contour detail, center location, size, and spiculation. Each version of segmentation is achieved by modifying the truth mask using a set of MISS-tool parameters governing the affine transformation, boundary spiculation, and Fourier descriptor as introduced in section~\ref{sec:fd}.  

Specifically, the 10 versions of segmentation are designed to have different combinations of the following characteristics: overall contour changes, location of center changes, segmentation size changes, and spiculations. These characteristics are considered to mimic segmentation errors empirically found in segmentations by human observers; in this sense, the MISS tool with 10 sets of parameter configurations serve as 10 simulated segmentors (virtual observers). The specific characteristics and parameter settings for each simulated segmentor are shown in Table~\ref{tab:obvs}. In general, the characteristics of 10 simulated segmentors are:
\begin{enumerate}
    \item moderate contour variation, size unchanged
    \item substantial contour variation, size unchanged
    \item substantial contour variation, center shifted, size unchanged
    \item moderate contour variation, over-segmentation
    \item moderate contour variation, under-segmentation
    \item moderate contour variation, center shifted, over-segmentation
    \item moderate contour variation, center shifted, under-segmentation
    \item moderate contour variation, 1-5 outward spiculations
    \item moderate contour variation, 1-5 inward spiculations
    \item moderate contour variation, 1-5 mixture spiculations
\end{enumerate}

\begin{table*}[]
\caption{Characteristics of the 10 simulated segmentors and the corresponding parameters of MISS synthesis methods. The \CheckmarkBold for the ``Similar contour?'' means the change of contour is moderate (Magnitude = 2) and $\times$ means a relatively substantial change (Magnitude = 8). For Fourier descriptor (FD), the ``Detail'' and ``Range'' show the percentages of FDs because the number of FDs vary with the number of pixels on contours.
The \CheckmarkBold for the ``Shifted?'' means the segmentation location is shifted (by 5-20 pixels on a random direction) and $\times$ means no shifting. ``Keep'' in ``Size'' means no size change; ``Oversize'' is 110\% resizing and ``Undersize'' is 85\% resizing by affine transformation. ``None'' in ``Spiculation'' means no spiculation shape is added to segmentation; ``1-5 outward'' means a random number (from 1-5) of outward spiculations are added; ``1-5 inward'' means a random number (from 1-5) of inward spiculations are added; ``1-5 mixture'' means a random number (from 1-5) of spiculations are added, for any spiculation, it could be outward or inward randomly. For each spiculation, the three parameters for generation are randomly selected from the ranges shown in the table.}
\label{tab:obvs}
\resizebox{\textwidth}{!}{\begin{tabular}{c|ccll|lll|ll|lll}
\hline
\multicolumn{1}{c|}{\multirow{3}{*}{Segmentor \#}} & \multicolumn{4}{c|}{Characteristics}   & \multicolumn{3}{c|}{Fourier Descriptor}    & \multicolumn{2}{c|}{Affine} & \multicolumn{3}{c}{Spiculation}   \\ \cline{2-13} 
\multicolumn{1}{c|}{}     & \scalebox{1}{\makecell{Similar \\ contour?}}  & Shifted? & Size  & Spiculation  & Detail & Range & Magnitude & Resize     & Shift  & Center    & Height   & Width  \\ \hline
1     & \CheckmarkBold    & \CheckmarkBold    & Keep & None     & 10\%   & 80\%  & 2     & 100\%  & 0  & NaN   & 0    & 0  \\
2     & \XSolid    & \CheckmarkBold    & Keep & None     & 10\%   & 80\%  & 8     & 100\%  & 0  & NaN   & 0    & 0  \\
3     & \XSolid    & \XSolid    & Keep & None     & 10\%   & 80\%  & 8     & 100\%  & {[}5,20{]}     & NaN   & 0    & 0  \\
4     & \CheckmarkBold    & \CheckmarkBold    & Oversize  & None     & 10\%   & 80\%  & 2     & 110\%  & 0  & NaN   & 0    & 0  \\
5     & \CheckmarkBold    & \CheckmarkBold    & Undersize & None     & 10\%   & 80\%  & 2     & 85\%   & 0  & NaN   & 0    & 0  \\
6     & \CheckmarkBold    & \XSolid    & Oversize  & None     & 10\%   & 80\%  & 2     & 110\%  & {[}5,20{]}     & NaN   & 0    & 0  \\
7     & \CheckmarkBold    & \XSolid    & Undersize & None     & 10\%   & 80\%  & 2     & 85\%   & {[}5,20{]}     & NaN   & 0    & 0  \\
8     & \CheckmarkBold    & \CheckmarkBold    & Keep & 1-5  outward & 10\%   & 80\%  & 2     & 100\%  & 0  & {[}0,360) & {[}3,25{]}   & {[}3,10{]} \\
9     & \CheckmarkBold    & \CheckmarkBold    & Keep & 1-5 inward & 10\%   & 80\%  & 2     & 100\%  & 0  & {[}0,360) & {[}-25,-3{]} & {[}3,10{]} \\
10    & \CheckmarkBold    & \CheckmarkBold    & Keep & 1-5 mixture  & 10\%   & 80\%  & 2     & 100\%  & 0  & {[}0,360) & {[}-25,25{]} & {[}3,10{]} \\ \hline
\end{tabular}}
\end{table*}

\begin{figure}[h]
    \centering
    \includegraphics[width=0.45\textwidth]{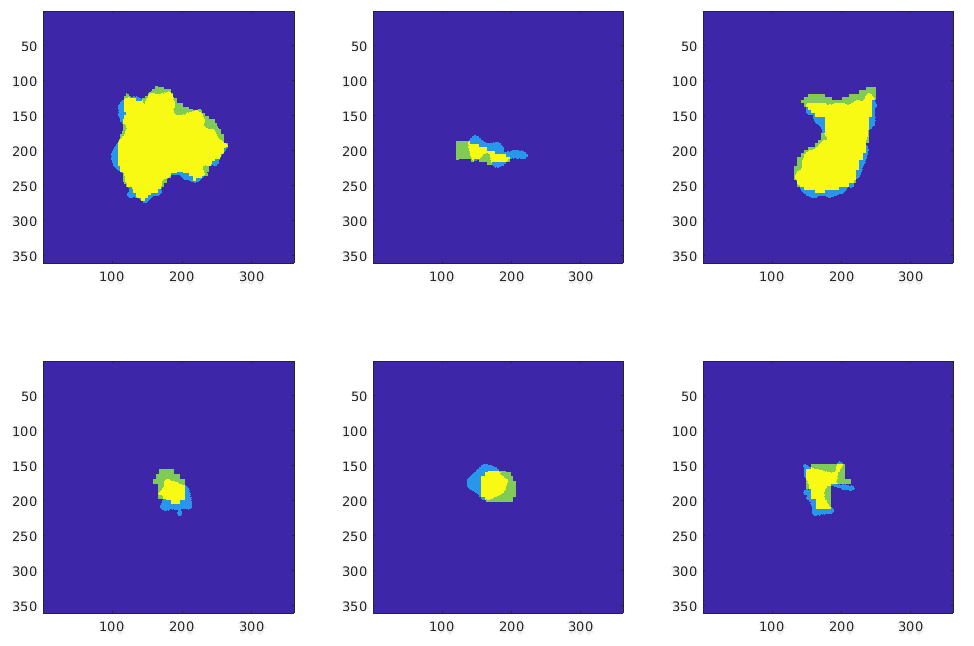}
    \caption{Six synthetic segmentations generated by the \#3 simulated segmentor: greatly contours different, center moved, size unchanged, and they are overlapped by according real segmentations. Yellow: True Positive; light blue: False Positive; green: False Negative; dark blue: True Negative.}
    \label{fig:syn_exps}
\end{figure}

In this study, 10 synthetic segmentations are generated by the 10 simulated segmentors for each truth mask of the 230 patients. In total, we have 2300 synthetic segmentations belonging to 10 groups and 230 pre-defined truth masks.  Figure~\ref{fig:syn_exps} shows examples of synthetic segmentations in the group \#3 (generated by the \#3 simulated segmentor), overlapped on the ground truth mask.

\subsection{Segmentation performance metrics}
Twenty metrics of segmentation performance are applied to evaluate the synthetic segmentations. The MSI is developed by Kim et al.~\cite{Kim2015Quantitative}, and others are from the literature summarized in Taha and Hanbury~\cite{Taha2015Metrics}. Table~\ref{tab:metric_list} lists these metrics.

\begin{table}[]
\caption{Overview of the metrics investigated in this study. The second column (Mono.) indicates the monotone embedding of these metrics; where greater(+) or smaller(-) value means the better performance.}
\label{tab:metric_list}
\centering
\begin{tabular}{lcl}
\hline
Symbol & Mono. & Metric   \\ \hline
DICE   & +     & Dice Coefficient     \\
JAC    & +     & Jaccard Coefficient  \\
MSI    & +     & Medical Similarity Index     \\
TPR    & +     & Sensitivity (Recall, true positive rate) \\
TNR    & +     & Specificity (True negative rate)     \\
FPR    & -     & Fallout (False positive rate)    \\
PPV    & +     & Precision (Positive predictive value)    \\
ACC    & +     & Accuracy     \\
AUC    & +     & Area under ROC Curve (One system state)  \\
VS     & +     & Volumetric Similarity Coefficient    \\
KAP    & +     & Cohen Kappa  \\
ARI    & +     & Adjusted Rand Index  \\
MI     & +     & Mutual Information   \\
VOI    & -     & Variation of Information     \\
GCE    & -     & Global Consistency Error     \\
ICC    & +     & Interclass Correlation   \\
PBD    & -     & Probabilistic Distance   \\
MHD    & -     & Mahanabolis Distance     \\
HD     & -     & Hausdorff Distance   \\
AVD    & -     & Average Hausdorff Distance   \\ \hline
\end{tabular}
\end{table}

\section{Experiments and results}
We use the synthetic segmentation data to investigate the properties of the 20 metrics:
\begin{itemize}
    \item Correlation of the ranking of segmentation methods among the metrics
    \item Scale/range of the metrics
    \item Effect of true negative regions on the metrics
\end{itemize}

\subsection{Ranking correlations of segmentation performance metrics}
We computed the 20 metrics (see Table~\ref{tab:metric_list}) for all synthetic segmentations: 10 segmentations for each truth mask of the 230 patients. For each metric, we rank the 10 segmentations on each patient from 1 to 10 with 1 referring to the best-performing segmentation and 10 the worst in terms of that metric. We summarize the ranking of the 10 segmentations over the 230 patients by taking the mode (i.e., the rank with the most number of patients).The results are shown in Table~\ref{tab:mode_ranks}.

\begin{table}[h]
\centering
\caption{Ranking of the 10 segmentations by 10 simulated segmentors using the 20 performance metrics (1 refers to best performance and 10 the worst). }
\label{tab:mode_ranks}
\begin{tabular}{cllllllllll}
\hline
\diagbox[width=7em]{\scalebox{0.95}{Metrics}}{\scalebox{0.95}{Segmentor}} & 1 & 2 & 3  & 4 & 5 & 6  & 7  & 8  & 9  & 10 \\ \hline
MSI & 1 & 2 & 10 & 4 & 2 & 8  & 9  & 7  & 6  & 6  \\
AVD & 1 & 2 & 10 & 4 & 2 & 8  & 9  & 7  & 7  & 6  \\
HD & 1 & 2 & 9  & 3 & 2 & 10 & 8  & 10 & 5  & 6  \\
DICE & 1 & 2 & 10 & 5 & 3 & 8  & 10 & 5  & 7  & 4  \\
JAC & 1 & 2 & 10 & 5 & 3 & 8  & 10 & 5  & 7  & 4  \\
KAP & 1 & 2 & 10 & 5 & 3 & 8  & 10 & 5  & 7  & 4  \\
ICC & 1 & 2 & 10 & 5 & 3 & 8  & 10 & 5  & 7  & 4  \\
PBD & 1 & 2 & 10 & 5 & 3 & 8  & 10 & 5  & 7  & 4  \\
ARI & 1 & 2 & 10 & 6 & 3 & 8  & 10 & 5  & 7  & 4  \\
ACC & 1 & 2 & 10 & 7 & 2 & 10 & 9  & 5  & 5  & 4  \\
GCE & 1 & 2 & 10 & 7 & 2 & 10 & 9  & 6  & 5  & 4  \\
VOI & 1 & 3 & 10 & 7 & 2 & 10 & 9  & 6  & 5  & 4  \\
VS & 1 & 3 & 3  & 9 & 7 & 9  & 7  & 4  & 10 & 5  \\
MI & 1 & 3 & 9  & 2 & 6 & 8  & 10 & 4  & 8  & 6  \\
TPR & 2 & 3 & 9  & 1 & 7 & 8  & 10 & 4  & 8  & 6  \\
AUC & 2 & 3 & 9  & 1 & 7 & 8  & 10 & 4  & 8  & 6  \\
MHD & 3 & 3 & 9  & 1 & 1 & 8  & 10 & 7  & 6  & 6  \\
PPV & 3 & 5 & 10 & 8 & 1 & 10 & 9  & 6  & 2  & 4  \\
TNR & 3 & 5 & 8  & 9 & 1 & 10 & 7  & 6  & 2  & 4  \\
FPR & 3 & 5 & 8  & 9 & 1 & 10 & 7  & 6  & 2  & 4  \\ \hline
\end{tabular}
\end{table}

We calculated the Pearson correlation coefficient of the rankings between all pairs of metrics . Specifically, the Pearson correlation coefficient between any two metrics (rows) in Table~\ref{tab:mode_ranks}, $R_i, R_j$, is:
\[
\rho_{i,j} = \frac{cov(R_i, R_j)}{\sigma_{R_i} \sigma_{R_j}},
\]
where $i,j\in \{1,2,3,\cdots, 20\}$ is the index to metrics; $R_i$ is the rank vector of length 10 for the $i^th$ metric, $cov(\cdot)$ is the covariance; $\sigma_{R_i}$ is the standard deviation of $R_i$. This method is similar to the Spearman's rank correlation coefficient~\cite{Myers2013statistical}.

\begin{figure*}[h]
    \centering
    \includegraphics[width=0.9\textwidth]{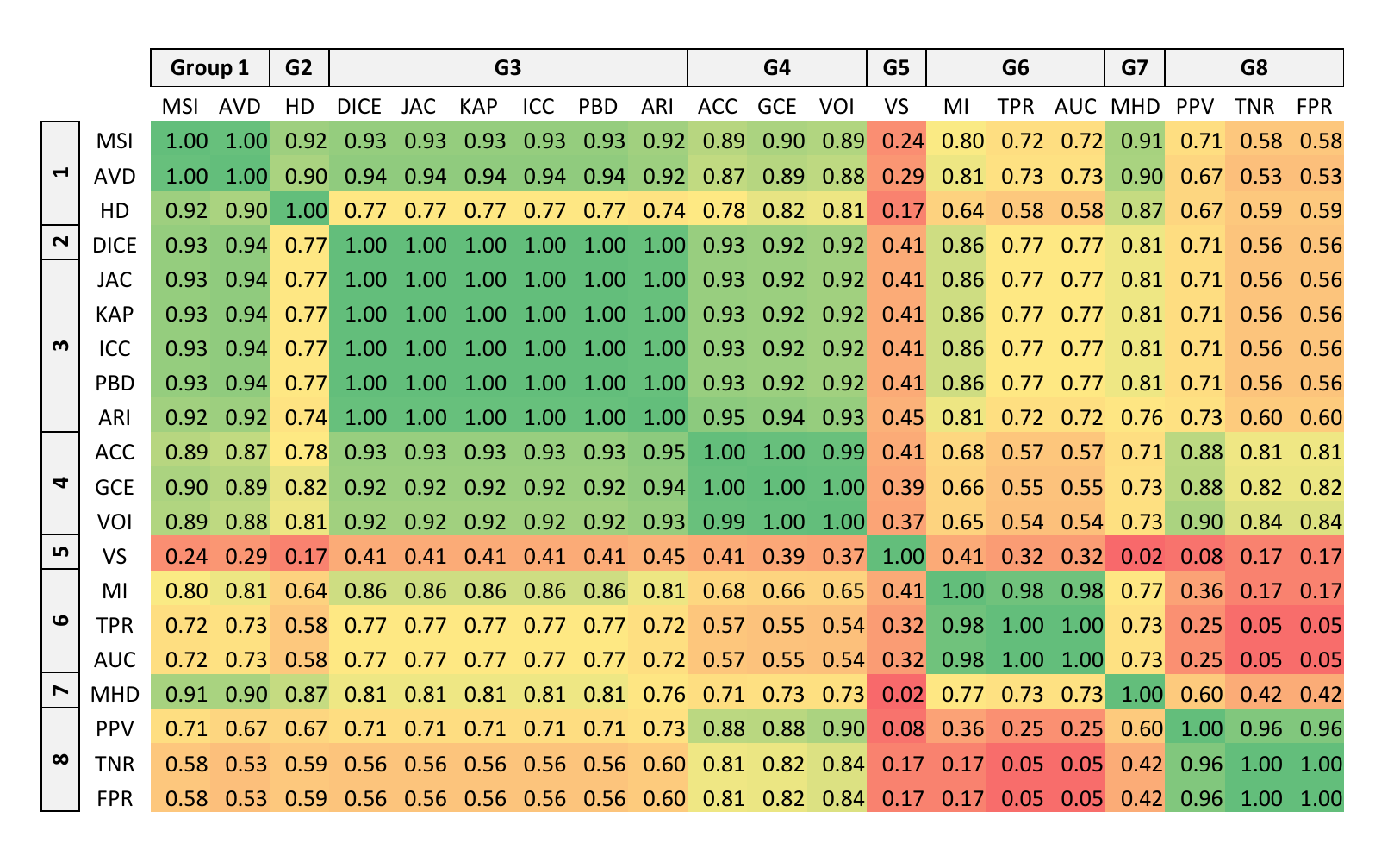}
    \caption{Ranking correlation coefficients between the metrics in Table~\ref{tab:mode_ranks}. High correlation shows in green and low correlation shows in red. Metrics have correlations close to 1 to each other are grouped; thus, eight groups are formed.}
    \label{fig:correlation}
\end{figure*}

Figure~\ref{fig:correlation} shows the rank correlation results. The metrics that are highly correlated are put into one group resulting 8 groups for the 20 metrics. 

\subsection{Ranges of metrics}
\begin{figure}[h]
    \centering
    \subfloat[\centering Best case ]{{\includegraphics[width=0.15\textwidth]{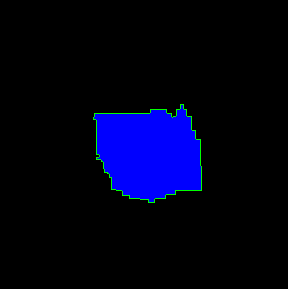} }}
    \subfloat[\centering Middle case ]{{\includegraphics[width=0.15\textwidth]{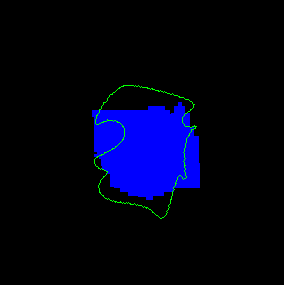} }}
    \subfloat[\centering Worst case ]{{\includegraphics[width=0.1516\textwidth]{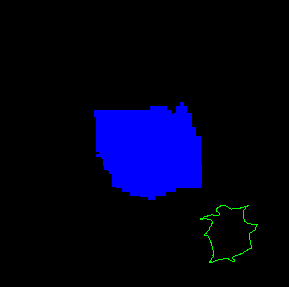} }}
    \caption{The three cases to test ranges of metrics. Blue area is the ground truth; green contours are segmentations. The segmentation for the best case is the same to the ground truth; for the worst case, there is no overlapping.}
    \label{fig:range}
\end{figure}

Since the range of values is an important property for metrics, and it is helpful for comparison different segmentation methods or interpretation of results. We synthesized three cases to illustrate this property. (Figure~\ref{fig:range}). Table~\ref{tab:range} shows the results of 20 metrics on the three cases.

\begin{table}[]
\centering
\caption{Results of 20 metrics on the three cases. The second column (Mono.) shows the monotone embedding of these metrics, where greater(+) or smaller(-) value means the better performance. \checkmark means metric values range in $[0,1]$; \textcircled{\checkmark} means metric values range in $[0,1]$ and the best and worst cases can reach 1 and 0; $\times$ means some values are out of $[0,1]$. The mark ``*'' in the row of PBD means that in theory, PBD is infinite for no overlapping cases but it is defined to be $-1$.}
\label{tab:range}
\begin{tabular}{lclllc}
\\ \hline
   Symbol  & Mono. & Best & Mid. & Worst & in [0,1] \\ \hline
DICE & +    & 1.000   & 0.742  & 0.000    & \textcircled{\checkmark}  \\
JAC  & +    & 1.000   & 0.590  & 0.000    & \textcircled{\checkmark}  \\
MSI  & +    & 1.000   & 0.491  & 0.000    & \textcircled{\checkmark}  \\
TPR  & +    & 1.000   & 0.710  & 0.000    & \textcircled{\checkmark}  \\
TNR  & +    & 1.000   & 0.977  & 0.904    & \checkmark  \\
FPR  & -    & 0.000   & 0.023  & 0.096    & \checkmark  \\
PPV  & +    & 1.000   & 0.778  & 0.000    & \textcircled{\checkmark}  \\
ACC  & +    & 1.000   & 0.949  & 0.885    & \checkmark  \\
AUC  & +    & 1.000   & 0.843  & 0.452    & \checkmark  \\
VS   & +    & 1.000   & 0.954  & 0.376    & \checkmark    \\
KAP  & +    & 1.000   & 0.714  & -0.036   & $\times$   \\
ARI  & +    & 1.000   & 0.670  & -0.032   & $\times$   \\
MI   & +    & 0.449   & 0.217  & 0.003    & \checkmark \\
VOI  & -    & 0.000   & 0.493  & 0.593    & \checkmark  \\
GCE  & -    & 0.000   & 0.092  & 0.136    & \checkmark  \\
ICC  & +    & 1.000   & 0.742  & 0.000    & \textcircled{\checkmark}  \\
PBD  & -    & 0.000   & 0.001  & -1.000   & $\times$*   \\
MHD  & -    & 0.000   & 0.136  & 2.817    & $\times$   \\
HD   & -    & 0.000   & 30.000 & 185.267  & $\times$   \\
AVD  & -    & 0.000   & 2.222  & 87.806   & $\times$  
\\ \hline
\end{tabular}
\end{table}

\subsection{True Negative area change}
\begin{figure*}[h]
    \centering
    \includegraphics[width=0.7\textwidth]{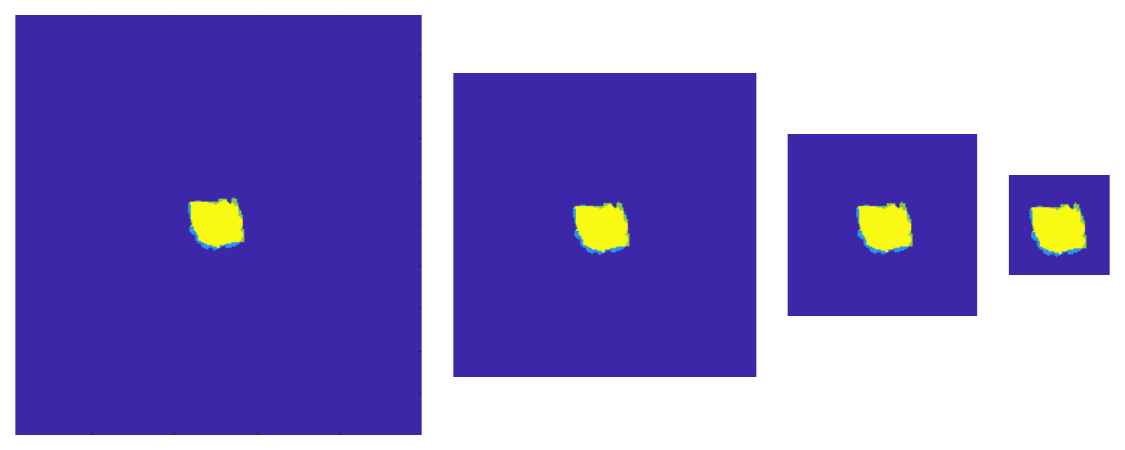}
    \caption{Examples for the change of True Negative (TN) area. Both segmentations and according ground truth keep the same size; only the background (TN) area is changed.
    Yellow: True Positive; light blue: False Positive; green: False Negative; dark blue: True Negative.}
    \label{fig:tn}
\end{figure*}

Figure~\ref{fig:tn} shows examples for the change of True Negative (TN) area. The areas of segmentations and according ground truth keep the same. Results in Table~\ref{tab:tn} shows the metrics: TNR, FPR, ACC, AUC, KAP, ARI, MI, VOI, GCE, and ICC change with the TN area.

\begin{table*}[]
\centering
\caption{Results of 20 metrics on 9 segmentations with different True Negative (TN) areas (pixels). As shown in Figure~\ref{fig:tn}, their TP, FN, and FP areas keep the same:(TP = 11217, FN = 926, FP = 1325; pixels). The bold font columns mean the metrics are changed with TN.}
\label{tab:tn}
\resizebox{\textwidth}{!}{
\begin{tabular}{l|llllllllllllllllllll}
\hline
\textbf{TN}  & DICE  & JAC   & MSI   & TPR   & \textbf{TNR}   & \textbf{FPR}   & PPV   & \textbf{ACC}   & \textbf{AUC}   & VS    & \textbf{KAP}   & \textbf{ARI}   & \textbf{MI}    & \textbf{VOI}   & \textbf{GCE}   & \textbf{ICC}   & PBD   & MHD   & HD  & AVD   \\ \hline
\textbf{3668}   & 0.909 & 0.833 & 0.737 & 0.894 & \textbf{0.798} & \textbf{0.202} & 0.924 & \textbf{0.869} & \textbf{0.846} & 0.984 & \textbf{0.674} & \textbf{0.526} & \textbf{0.320} & \textbf{1.069} & \textbf{0.238} & \textbf{0.908} & 0.000 & 0.241 & 14.318 & 0.350 \\
\textbf{9032}   & 0.909 & 0.833 & 0.737 & 0.894 & \textbf{0.907} & \textbf{0.093} & 0.924 & \textbf{0.900} & \textbf{0.901} & 0.984 & \textbf{0.798} & \textbf{0.640} & \textbf{0.527} & \textbf{0.933} & \textbf{0.189} & \textbf{0.908} & 0.000 & 0.241 & 14.318 & 0.350 \\
\textbf{26532}  & 0.909 & 0.833 & 0.737 & 0.894 & \textbf{0.966} & \textbf{0.034} & 0.924 & \textbf{0.944} & \textbf{0.930} & 0.984 & \textbf{0.868} & \textbf{0.783} & \textbf{0.587} & \textbf{0.609} & \textbf{0.108} & \textbf{0.909} & 0.000 & 0.241 & 14.318 & 0.350 \\
\textbf{49032}  & 0.909 & 0.833 & 0.737 & 0.894 & \textbf{0.981} & \textbf{0.019} & 0.924 & \textbf{0.964} & \textbf{0.938} & 0.984 & \textbf{0.886} & \textbf{0.840} & \textbf{0.506} & \textbf{0.421} & \textbf{0.070} & \textbf{0.909} & 0.000 & 0.241 & 14.318 & 0.350 \\
\textbf{76532}  & 0.909 & 0.833 & 0.737 & 0.894 & \textbf{0.988} & \textbf{0.012} & 0.924 & \textbf{0.975} & \textbf{0.941} & 0.984 & \textbf{0.894} & \textbf{0.865} & \textbf{0.423} & \textbf{0.308} & \textbf{0.048} & \textbf{0.909} & 0.000 & 0.241 & 14.318 & 0.350 \\
\textbf{116132} & 0.909 & 0.833 & 0.737 & 0.894 & \textbf{0.992} & \textbf{0.008} & 0.924 & \textbf{0.983} & \textbf{0.943} & 0.984 & \textbf{0.899} & \textbf{0.880} & \textbf{0.341} & \textbf{0.224} & \textbf{0.034} & \textbf{0.909} & 0.000 & 0.241 & 14.318 & 0.350 \\
\textbf{236532} & 0.909 & 0.833 & 0.737 & 0.894 & \textbf{0.996} & \textbf{0.004} & 0.924 & \textbf{0.991} & \textbf{0.945} & 0.984 & \textbf{0.904} & \textbf{0.895} & \textbf{0.221} & \textbf{0.126} & \textbf{0.017} & \textbf{0.909} & 0.000 & 0.241 & 14.318 & 0.350 \\
\textbf{626532} & 0.909 & 0.833 & 0.737 & 0.894 & \textbf{0.999} & \textbf{0.001} & 0.924 & \textbf{0.996} & \textbf{0.946} & 0.984 & \textbf{0.907} & \textbf{0.904} & \textbf{0.110} & \textbf{0.054} & \textbf{0.007} & \textbf{0.909} & 0.000 & 0.241 & 14.318 & 0.350 \\
\textbf{986532} & 0.909 & 0.833 & 0.737 & 0.894 & \textbf{0.999} & \textbf{0.001} & 0.924 & \textbf{0.998} & \textbf{0.947} & 0.984 & \textbf{0.908} & \textbf{0.906} & \textbf{0.078} & \textbf{0.036} & \textbf{0.004} & \textbf{0.909} & 0.000 & 0.241 & 14.318 & 0.350 \\ \hline
\end{tabular}
}
\end{table*}

\section{Discussions}
Previoulsy, Taha and Hanbury~\cite{Taha2015Metrics} studied the correlation between sixteen performance metrics using 4833 segmentations. In our study, we examined the correlation of metrics in ranking 10 simulated simulated segmentors (\textit{i.e.}, segmentation methods). The use of synthetic segmentation in our method allows systematic analysis of different types of segmentation errors and how performance metrics respond to different errors.

Specifically, by combining the ranking results in Table~\ref{tab:mode_ranks} and the grouping of metrics based on the ranking correlations(Figure~\ref{fig:correlation}), we can summarize further to show the best and worst segmentation performance among the 10 simulated segmentors in terms of each group of metrics, as shown in Table~\ref{tab:groups}. 

\begin{table*}[h]
\centering
\caption{Ranking the segmentation performance of the 10 simulated segmentors by the 8 groups of metrics}
\label{tab:groups}
\begin{tabular}{c|cc|l|llllllllll}
\hline
\multicolumn{1}{c|}{Group \#} & \makecell{Best \\ Segmentor} & \makecell{Worst \\ Segmentor} & \diagbox[width=7em]{\scalebox{0.95}{Metrics}}{\scalebox{0.95}{Segmentor}} & 1 & 2 & 3  & 4 & 5 & 6  & 7  & 8  & 9  & 10 \\ \hline
\multirow{2}{*}{1}    & \multirow{2}{*}{1}    & \multirow{2}{*}{3}    & MSI    & 1 & 2 & 10 & 4 & 2 & 8  & 9  & 7  & 6  & 6  \\
  &    &    & AVD    & 1 & 2 & 10 & 4 & 2 & 8  & 9  & 7  & 7  & 6  \\ \hline
2   & 1  & 6, 8    & HD     & 1 & 2 & 9  & 3 & 2 & 10 & 8  & 10 & 5  & 6  \\ \hline
\multirow{6}{*}{3}    & \multirow{6}{*}{1}    & \multirow{6}{*}{3, 7} & DICE  & 1 & 2 & 10 & 5 & 3 & 8  & 10 & 5  & 7  & 4  \\
  &    &    & JAC    & 1 & 2 & 10 & 5 & 3 & 8  & 10 & 5  & 7  & 4  \\
  &    &    & KAP    & 1 & 2 & 10 & 5 & 3 & 8  & 10 & 5  & 7  & 4  \\
  &    &    & ICC    & 1 & 2 & 10 & 5 & 3 & 8  & 10 & 5  & 7  & 4  \\
  &    &    & PBD    & 1 & 2 & 10 & 5 & 3 & 8  & 10 & 5  & 7  & 4  \\
  &    &    & ARI    & 1 & 2 & 10 & 6 & 3 & 8  & 10 & 5  & 7  & 4  \\ \hline
\multirow{3}{*}{4}    & \multirow{3}{*}{1}    & \multirow{3}{*}{3, 6} & ACC    & 1 & 2 & 10 & 7 & 2 & 10 & 9  & 5  & 5  & 4  \\
  &    &    & GCE    & 1 & 2 & 10 & 7 & 2 & 10 & 9  & 6  & 5  & 4  \\
  &    &    & VOI    & 1 & 3 & 10 & 7 & 2 & 10 & 9  & 6  & 5  & 4  \\ \hline
5   & 1  & 9  & VS     & 1 & 3 & 3  & 9 & 7 & 9  & 7  & 4  & 10 & 5  \\ \hline
\multirow{3}{*}{6}    & \multirow{3}{*}{1, 4} & \multirow{3}{*}{7}    & MI     & 1 & 3 & 9  & 2 & 6 & 8  & 10 & 4  & 8  & 6  \\
  &    &    & TPR    & 2 & 3 & 9  & 1 & 7 & 8  & 10 & 4  & 8  & 6  \\
  &    &    & AUC    & 2 & 3 & 9  & 1 & 7 & 8  & 10 & 4  & 8  & 6  \\ \hline
7   & 4, 5    & 7  & MHD    & 3 & 3 & 9  & 1 & 1 & 8  & 10 & 7  & 6  & 6  \\ \hline
\multirow{3}{*}{8}    & \multirow{3}{*}{5}    & \multirow{3}{*}{6}    & PPV    & 3 & 5 & 10 & 8 & 1 & 10 & 9  & 6  & 2  & 4  \\
  &    &    & TNR    & 3 & 5 & 8  & 9 & 1 & 10 & 7  & 6  & 2  & 4  \\
  &    &    & FPR    & 3 & 5 & 8  & 9 & 1 & 10 & 7  & 6  & 2  & 4  \\ \hline
\end{tabular}
\end{table*}

By referring to the definitions of these simulated segmentors (Table~\ref{tab:obvs}), we can understand the properties of these segmentation metrics (ranking \textit{vs.} characteristics of segmentors) and provide selection suggestions to users. 
For example, in Group \#1, the two metrics: MSI and AVD give the best ranking to Segmentor \#1 and the worst ranking to Segmentor \#3. By definition, there is no surprise that Segmentor \#1 is the best; MSI and AVD consider Segmentor \#3 is the worst and Segmentor \#3 has big change of contour and shifted center that lead to large boundary distances. For another example in Group \#8, the three metrics give the best ranking to Segmentor \#5 but its known to be an under-segmentation by design, which reveals the metrics in Group \#8 may not be sensitive to under-segmentation.

In future work, we will extend our current work to a more comprehensive investigation of the properties of segmentation performance metrics. We will synthesize segmentation errors represented by satellite (outlier) structures, i.e., there are more than one segmentation components in an image. We will define more simulated segmentors by different combinations of characteristics (segmentation errors) and parameters to generate these characteristics. Also, the MISS tool can be used to investigate the properties of truthing methods. By further analyzing the intrinsic properties of these metrics and categorizing the segmentation errors, we are working toward the goal of developing a decision-tree type of tool for assisting selection of segmentation performance metrics.

\section{Conclusions}
We developed a tool to generate synthetic segmentations with controllable characteristics (segmentation errors) based on pre-defined segmentations from real medical images. Our tool contains a set of user-configurable parameters that can modify the contours to emulate different types of segmentation errors.
The pre-defined ground truth and synthetic segmentations can be used to inform selection of performance metrics for evaluation of medical image segmentation algorithms.

We simulated 10 simulated segmentors with various segmentation errors and used our segmentation synthesis tool to emulate the segmentation results from the 10 simulated segmentors. We generated synthetic segmentations using 230 truth masks based on patient images in the GLIS-RT database, computed 20 performance metrics on the synthetic segmentations and compared their rankings of the 10 simulated segmentors. Our experiments, despite preliminary, demonstrate the feasibility of using our tool to investigate properties of segmentation performance metrics including their ability to capture specific types of segmentation errors and inform metrics selection decisions.

\section{Disclaimers}
The mention of commercial products, their sources, or their use in connection with material reported herein is not to be construed as either an actual or implied endorsement of such products by the Department of Health and Human Services. This is a contribution of the U.S. Food and Drug Administration and is not subject to copyright.

\section{Acknowledgment}
This work was supported in part by the FDA Office of Chief Scientist. This project was supported in part by an appointment to the ORISE Research Participation Program at the Center for Devices and Radiological Health, U.S. Food and Drug Administration, administered by the Oak Ridge Institute for Science and Education through an interagency agreement between the U.S. Department of Energy and FDA/CDRH.


\bibliography{ref}


\end{document}